\documentclass[epj]{svjour}
\usepackage[english]{babel}
\usepackage{lineno}
\usepackage{amssymb}
\usepackage{amsmath}
\usepackage{multirow}
\usepackage{latexsym}
\usepackage{textcomp}
\usepackage{amsmath}
\usepackage[numbers,sort&compress]{natbib}
\usepackage{graphicx}% Include figure files
\usepackage{dcolumn}% Align table columns on decimal point
\usepackage[shortcuts]{extdash}
\begin{document}

%\preprint{APS/123-QED}

\title{Trapped-ion decay spectroscopy towards the determination of
ground-state components of double-beta decay matrix elements}
\titlerunning{Trapped-ion decay spectroscopy for double-beta decay studies}
% Force line breaks with \\
%\author{TITAN-EC collaboration et al}
%\author{T. Brunner\inst{1}}
\author{T. Brunner \inst{1\and 2}\thanks{\emph{Present address:} Stanford University, Stanford, CA, 94305, USA\fnmsep\email{tbrunner@stanford.edu}}\and A. Lapierre\inst{1}\thanks{\emph{Present address:} NSCL, Michigan State University, East Lansing, MI, 48824, USA} \and C. Andreoiu\inst{3} \and M. Brodeur\inst{4} \and P. Delheji\inst{1} \and S. Ettenauer\inst{1\and 5}\thanks{\emph{Present address:} Department of Physics, Harvard University, Cambridge, Massachusetts 02138, USA} \and D. Frekers\inst{6} \and A.T. Gallant\inst{1\and 5} \and R. Gernh\"auser\inst{2} \and A. Grossheim\inst{1} \and R. Kr\"ucken\inst{2}\thanks{\emph{Present address:} TRIUMF, Vancouver, BC, V6T 2A3, Canada} \and A. Lennarz\inst{1\and 6} \and D. Lunney\inst{7} \and D. M\"ucher\inst{2} \and R. Ringle\inst{1}\thanks{\emph{Present address:} NSCL, Michigan State University, East Lansing, MI, 48824, USA} \and M.C. Simon\inst{1}\thanks{\emph{Present address:} Stefan Meyer Institute for subatomic Physics, Vienna, Austria} \and V.V. Simon\inst{1\and 8\and 9}\thanks{\emph{Present address:} Helmholtz Institute Mainz, Germany} \and S.K.L. Sjue\inst{1}\thanks{\emph{Present address:} Los Alamos National Laboratory, Los Alamos, NM, 87544, USA} \and K. Zuber\inst{10} \and J. Dilling\inst{1,5}}
%%%%%%%%
\institute{TRIUMF, Vancouver, BC, V6T 2A3, Canada \and
 					Physik Department E12, Technische Universit\"at M\"unchen, D--85748 Garching, Germany \and
					Department of Chemistry, Simon Fraser University, Burnaby, BC, V5A 1S6, Canada \and
					Department of Physics, University of Notre Dame, Notre Dame, IN, 46556 US \and
 					Department of Physics and Astronomy, University of British Columbia, Vancouver, BC, V6T 1Z1, Canada\and
 					Westf\"alische Wilhelms-Universit\"at M\"unster, D--48149 M\"unster, Germany\and
 					CSNSM-IN2P3-CNRS, Universit\'e de Paris Sud, 91405 Orsay, France\and
 					Fakul\"at f\"ur Physik und Astronomie, Ruprecht-Karls-Universit\"at Heidelberg, D--69120 Heidelberg, Germany\and
 					Max-Planck-Institut f\"ur Kernphysik, D--69117 Heidelberg, Germany\and
 					Institut f\"ur Kern- und Teilchenphysik, Technische Universit\"at Dresden, D--01069 Dresden, Germany}
%%
%\author{et al.}
%\affiliation{TRIUMF, 4004 Wesbrook Mall, Vancouver, V6T 2A3, Canada}
%\affiliation{Department of Physics and Astronomy, University of British Columbia, 6224~Agriculture Road, Vancouver, BC, V6T 1Z1, Canada}
%


%\affiliation{
%\affiliation{
%\affiliation{%


%%\date{\today}% It is always \today, today,
%%             %  but any date may be explicitly specified
%\linenumbers
\abstract{
A new technique has been developed at TRIUMF's TITAN facility to perform in-trap decay
spectroscopy. The aim of this technique is to eventually measure weak electron capture branching ratios (ECBRs) and by this to consequently determine GT matrix elements of $\beta\beta$ decaying nuclei. These branching ratios provide important input to the theoretical description of these decays. The feasibility and power of the technique is demonstrated by measuring the ECBR of $^{124}$Cs. 
\PACS{
			{37.10.Ty}{} \and
			{29.30.Kv} \and
			{23.40.-s} \and
			{14.60.Lm} \and
			{14.60.St}{}
			}% PACS, the Physics and Astronomy
%                             % Classification Scheme.
\keywords{Nuclear matrix element, neutrinoless double-beta decay, Penning trap,
in-trap decay spectroscopy}%Use showkeys class option if keyword
}                              %display desired
\maketitle
\date
%\hyphenation{Stu-dien-stif-tung, Deut-schen}
%\tableofcontents
\section{The $\beta\beta$ decay matrix element}
Measuring the neutrino properties poses great challenges in modern physics. Despite numerous efforts, the character of the neutrino still remains a mystery, i.e., whether it is a Dirac or Majorana particle \cite{Hax84,Doi85}. A powerful approach to shed light on this question is through $\beta\beta$-decay experiments \cite{Hir05,Avi08,Zub12}. Double-beta decay is a rare, second-order weak process, which is expected to happen in at least two modes, the zero-neutrino ($0\nu\beta\beta$) and the two-neutrino ($2\nu\beta\beta$) mode. 
%All $\beta\beta$-decay candidates are even-even nuclei, hence the decay is a $0^+\rightarrow0^+$ transition. 
The $2\nu\beta\beta$ decay has been observed in various isotopes with half-lives greater than $10^{18}$ years \cite{Bar10}. The $0\nu\beta\beta$ mode is forbidden in the Standard Model of particle physics as it violates lepton-number conservation. Moreover, the existence of the neutrinoless mode uniquely establishes the Majorana character of the neutrino \cite{Sch82,Tak84}, and if the mass term is the leading contribution, one can deduce the effective Majorana neutrino mass $\langle m_{\nu}\rangle$ from the half-life of the decay:
\begin{equation}
\left(T^{0\nu}_{1/2}\right)^{-1}=G^{0\nu}\left(Q,Z\right)\left|M^{0\nu}\right|^2\langle
m_{\nu}\rangle^2.
\label{eq:halflife}
\end{equation}
Here, $G^{0\nu}\left(Q,Z\right)$ is the phase-space factor and $M^{0\nu}$ is the
nuclear matrix element (NME).
% with the latter being entirely based on theoretical calculations. 
Currently, several models provide values for $M^{0\nu}$ (see for example \cite{Hax84,Geh07,Eng09,Bar09,Rod10,Rat12,Cha09,Men11}). 
%However, when the results obtained within these different frameworks are compared with one another (or with each other), deviations arise. 
In order to extract a value of $m_{\nu}$, $M^{0\nu}$ must be known with reasonable precision.
%In oder to extract a meaningful value for $m_{\nu}$ one needs to know $M^{0\nu}$ with an uncertainty of less than 50\%.% \cite{Aki97}.
%\begin{figure}
        %\centering
                %\includegraphics[width=.48\textwidth]{NME}
        %\caption{\label{fig:NME}(color online) $M^{0\nu}$ calculated in the Interacting Boson
%Model $\left(\boxdot\right)$ \cite{Bar09} and within the Shell Model $\left(\odot\right)$\cite{Cau08}. Average $\left\langle M^{0\nu}\right\rangle$ calculated within the renormalized (R)QRPA applying different treatments of short-range correlations and averaging over different parameter sets (Jastrov $\left(\bigtriangleup\right)$ and Brueckner$\left(\bigtriangledown\right)$ from \cite{Sim09}). The error bars were obtained by varying the initial parameter set.}
%\end{figure}
%%%%%%%%%%%%%% Jan10, 2013 -->
%The present situation is reflected in Fig.~\ref{fig:NME} displaying $M^{0\nu}$ calculated within different frameworks. Calculations disagree sometimes even within the same framework when employed by different groups. 
%%%%%%%%%%%%%%%%%%%
Frequently, theoretical approaches use the $2\nu\beta\beta$ decay to test their models (see for example \cite{Sim04}). 
%In these decays Fermi transitions are forbidden or at least strongly suppressed \cite{Hax84}, hence the transition matrix element is dominated by the Gamow-Teller contribution. 
%In this case, the $2\nu\beta\beta$ transition can be described as a sequence of two single $\beta$ decays via $1^+$ states in the intermediate odd-odd nucleus. The $2\nu\beta\beta$ matrix element is then the sum over all possible intermediate states and is connected to the decay rate by a phase-space factor. 
Recent experiments have shown that the $2\nu$ NME is rather sensitive to the ground-state properties of the nuclear wave function \cite{Thi12a,Thi12,Eji09,Eji12}. An experimental access to probe these ground-state nuclear wave functions and thus part of the Gamow-Teller matrix element is through spectroscopic measurements (see Ref. \cite{Fre07} and references therein). They can be obtained from measurements of the branching ratios of the intermediate nucleus, i.e., the electron capture (EC) and $\beta$ branching ratios (BRs) of the ground-state decay of the intermediate nucleus. %This measurement directly probes the ground state wave function and thus part of the Gamow-Teller matrix element (see Ref. \cite{Fre07} and references therein). 
%%%%%%%%%%%%%%%%%% Jan10, 2013 -->
%In special cases, such as $^{100}$Mo, a so-called single-state dominance (SSD) is suggested \cite{Mor09}, where the transition via the lowest $1^+$ state is the main contributor to $M^{2\nu}$ \cite{Aba84}. In the case of $^{116}$Cd a single-state dominance is likely, but experimental uncertainties are not sufficiently small to allow a definitive claim \cite{Dom05,Mor09}. As a consequence, in SSD nuclei one determines the main contribution to $M^{2\nu}$ by measuring the EC and $\beta$ branches of the intermediate nuclei. 
%%%%%%%%%%%%%%
For all intermediate nuclei in $\beta\beta$ decays the ECBRs are small
%, i.e., of the order of $10^{-5}$ (for example in the case of $^{100}$Tc), 
 and the EC signature is the emission of X-rays by the daughter atom. This poses a formidable experimental challenge for ECBR measurements. In addition, the short-lived intermediate nucleus has to be produced at a radioactive isotope facility, which almost always delivers isobaric contamination.

In previous ECBR measurements such as the one of $^{100}$Tc, the ECBR has been measured by the implantation of radioactive $^{100}$Tc onto a tape \cite{Gar93}. This tape was then moved in front of several X-ray and $\beta$-particle detectors. The difficulties of this method arise from contaminations in the sample, the $\beta$-particle background from dominating $\beta$ branches, and possible X-ray attenuation in the
implantation material. Improvements of this measurement were accomplished by applying the technique of trap-assisted decay spectroscopy at Jyv\"askyl\"a \cite{Sju09}. Here, the sample was isobarically purified in a Penning trap by means of a mass-selective buffer-gas cooling technique \cite{Sav91}. The purified sample was then implanted onto an Al foil in front of a X-ray detector. The Al foil was embedded into a plastic scintillator used to veto $\beta$ particles. This technique improved the uncertainty of the previous ECBR measurement of $^{100}$Tc by a factor of two \cite{Sju09}. 
However, it still suffers from a large background due to secondary radiation and X-ray attenuation by the implantation material. In order to overcome these drawbacks a novel technique has been proposed \cite{Fre07} and developed at TITAN using an open-access Penning trap to
perform in-trap decay spectroscopy. In this scheme the strong magnetic field of the trap guides $\beta$ particles away from the X-ray detectors and thereby reduces their contribution to the X-ray spectrum. Furthermore, established ion-trap techniques available at TITAN allow for isobaric purification of the sample in the future. Ion traps are nowadays well established tools for precision experiments at radioactive beam facilities \cite{Bla13}, however, this in-trap application is unique to the TITAN set-up.
%The combination of these techniques offers a significant improvement over previous techniques to measure ECBRs. 
In this paper the feasibility of this technique is demonstrated by measuring the ECBR of $^{124}$Cs. 
%\begin{figure}
%        \centering
%                \includegraphics[width=.4\textwidth]{decay-scheme}
%        \caption{\label{fig:decay-scheme}Decay scheme of a $\beta\beta$ decay via the
%transition nucleus (left) and the decay scheme or a $\beta^+$ and EC decay
%(right).}
%\end{figure}
%%%%%%%%%%%%%%%%%%%%%%%%%%%%%%%%%%%%%%%%%%% TITAN-EC setup
\begin{figure}
        \centering
                \includegraphics[width=.47\textwidth]{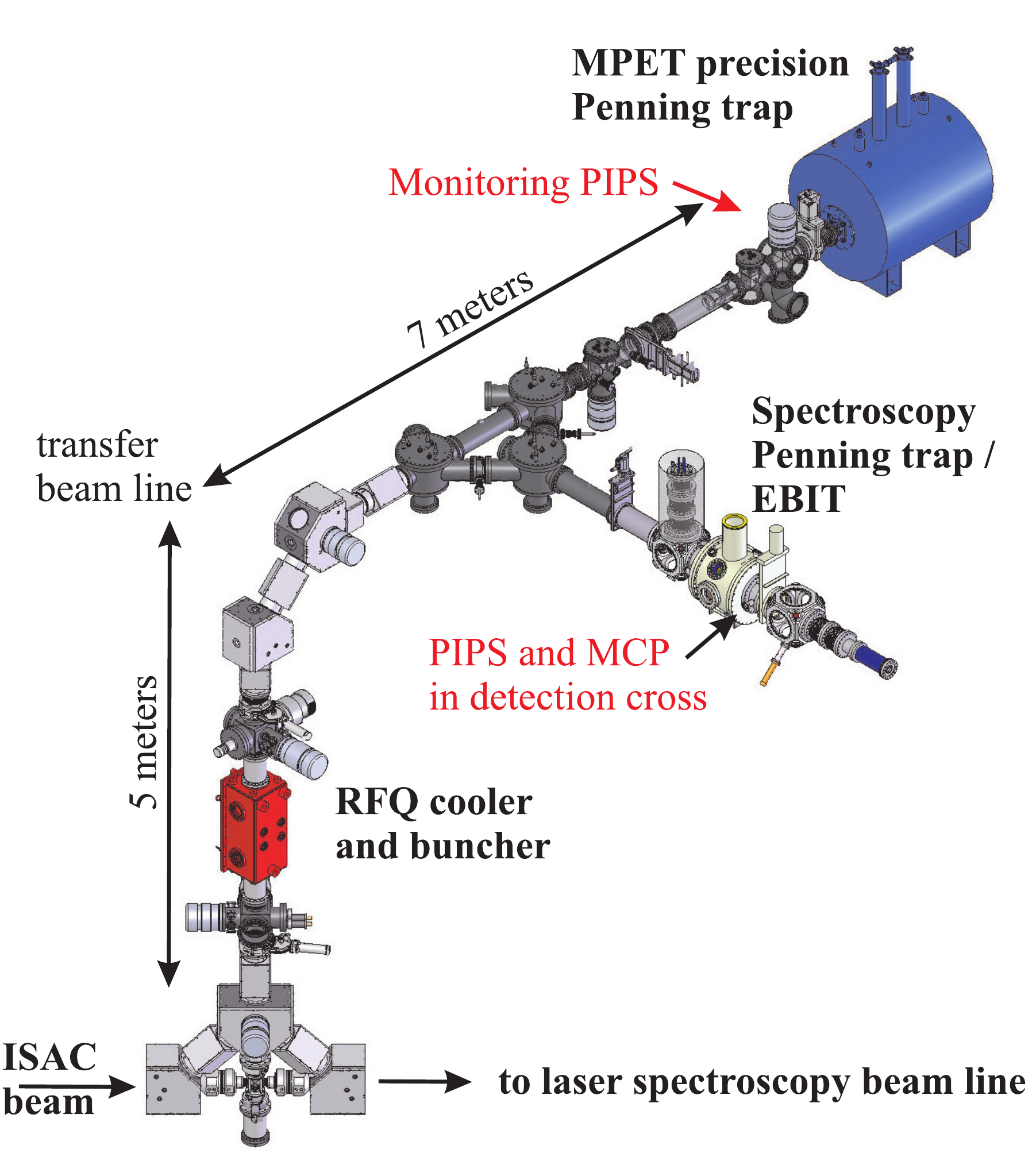}
        \caption{\label{fig:TITAN-diagnostics}(color online) Engineering rendering of the TITAN facility showing RFQ, MPET, EBIT as well as the positions of the PIPS detectors.}
\end{figure}
\section{TITAN-EC Setup}
%The in-trap decay spectroscopy technique employing a Penning trap has been developed and tested at one of TITAN's ion traps. 
TITAN \cite{Dil03,Dil06} is TRIUMF's Ion Trap
facility for Atomic and Nuclear science which currently consists of the Radio-Frequency Quadrupole (RFQ)
\cite{Smi06,Bru10b}, the Electron Beam Ion Trap (EBIT) \cite{Sik05,Lap10} and the mass
Measurement Penning Trap (MPET) \cite{Bro11c}. An engineering model of the experimental setup is
given in Fig.~\ref{fig:TITAN-diagnostics}.
\begin{figure}
        \centering
                \includegraphics[width=.48\textwidth]{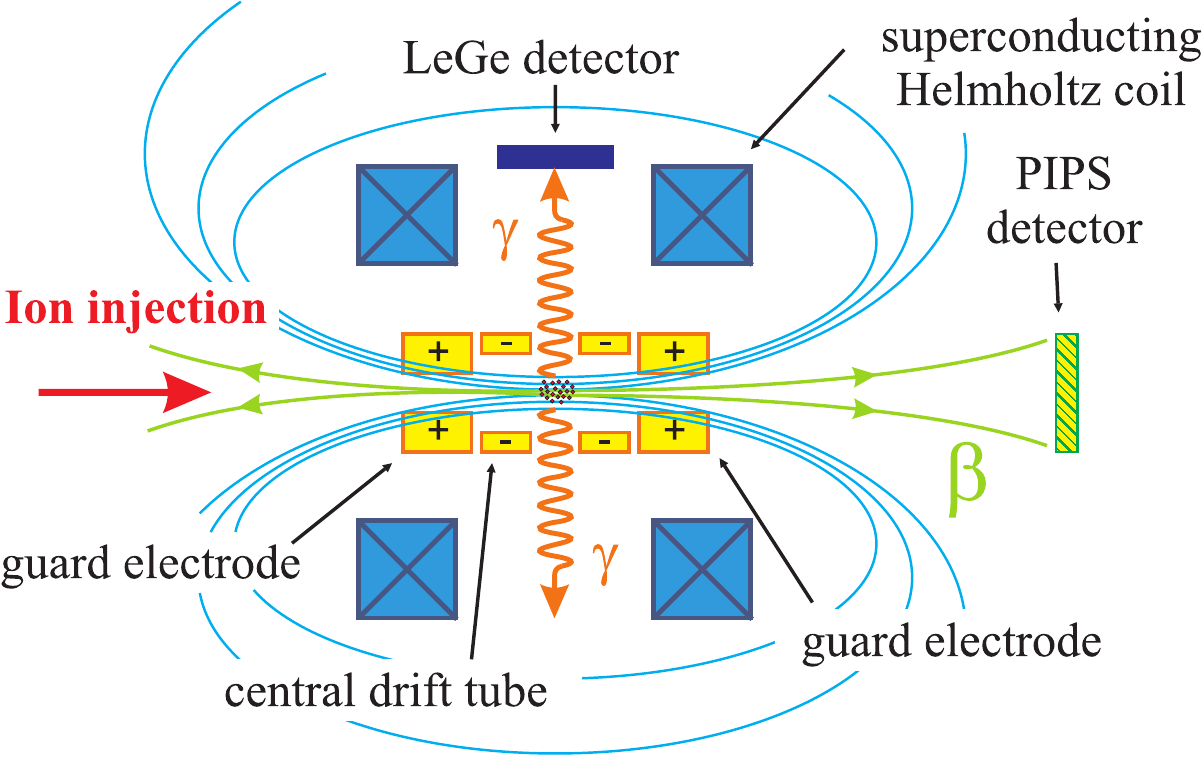}
        \caption{\label{fig:TITANEC}(color online) Schematic illustration of the in-trap decay spectroscopy Penning trap at TITAN.  After injection, radioactive ions are trapped for up to 50\,ms. While stored inside the trap, their electron-capture decays are monitored by a LeGe detector installed on a radial view port perpendicular to the ion beam line. Electrons resulting from $\beta$ decays are guided out of the trap-center region towards both ends by the trap's strong magnetic field. Beta particles are detected with a PIPS detector at one end.}
\end{figure}

For this ECBR measurement at TITAN, the EBIT is operated as a spectroscopy-Penning trap without the electron beam. During in-trap decay spectroscopy measurements radioactive ions are stored at the center of the trap. Gamma- and X-rays following an EC event are emitted isotropically, while $\beta$
particles are guided out of the trap along the axis of the magnetic field (up to 5\,T in the present experiment, see Fig.\,\ref{fig:TITANEC}). These $\beta$ particles are detected by a Passivated Implanted Planar Silicon (PIPS) detector placed at one side of the trap \cite{Bru11} labeled PIPS in Fig.\,\ref{fig:TITAN-diagnostics} and Fig.\,\ref{fig:TITANEC}. The radial confinement provided by the magnetic field prevents electrons from reaching the X-ray detector. This spatial separation between X-ray and $\beta$-particle detection constitutes the significant advantage of this technique compared to conventional
ones. A schematic illustration of the setup of the spectroscopy trap is presented in Fig.~\ref{fig:TITANEC}.

The radially confining magnetic field of the trap is generated by a pair of superconducting coils arranged in a Helmholtz-like configuration \cite{Lap10}. This configuration allows for visible access to trapped ions owing to radial openings in the magnet assembly and the central trap electrode. This electrode is eight-fold segmented with slit apertures
between each segment. %A photo of the central trap electrode is presented in Fig.~\ref{fig:trap-center}. 
In total, seven of these view ports towards the trap center are available for in-trap decay spectroscopy \cite{Lap10}. The measurements presented here were obtained with only one low-energy planar Ge detector (LeGe, Canberra model GUL 0110P, \o 11.3\,mm, 10\,mm thick) used for $\gamma$- and X-ray detection. It was installed inside the vacuum vessel and positioned between the coils at a distance of $\sim$10\,cm from the trap center.
%\begin{figure}
%        \centering
%                \includegraphics[width=.48\textwidth]{trap-center-pic}
%        \caption{\label{fig:trap-center}(color online) Photo of the trap
%center assembly. Ceramic rings electrically insulate the copper electrodes while hidden sapphire discs provide thermal contact \cite{Lap10}. Apertures in the central electrode allow a view to the trap center.}
%\end{figure}

\begin{figure}
	\centering
		\includegraphics[width=.48\textwidth]{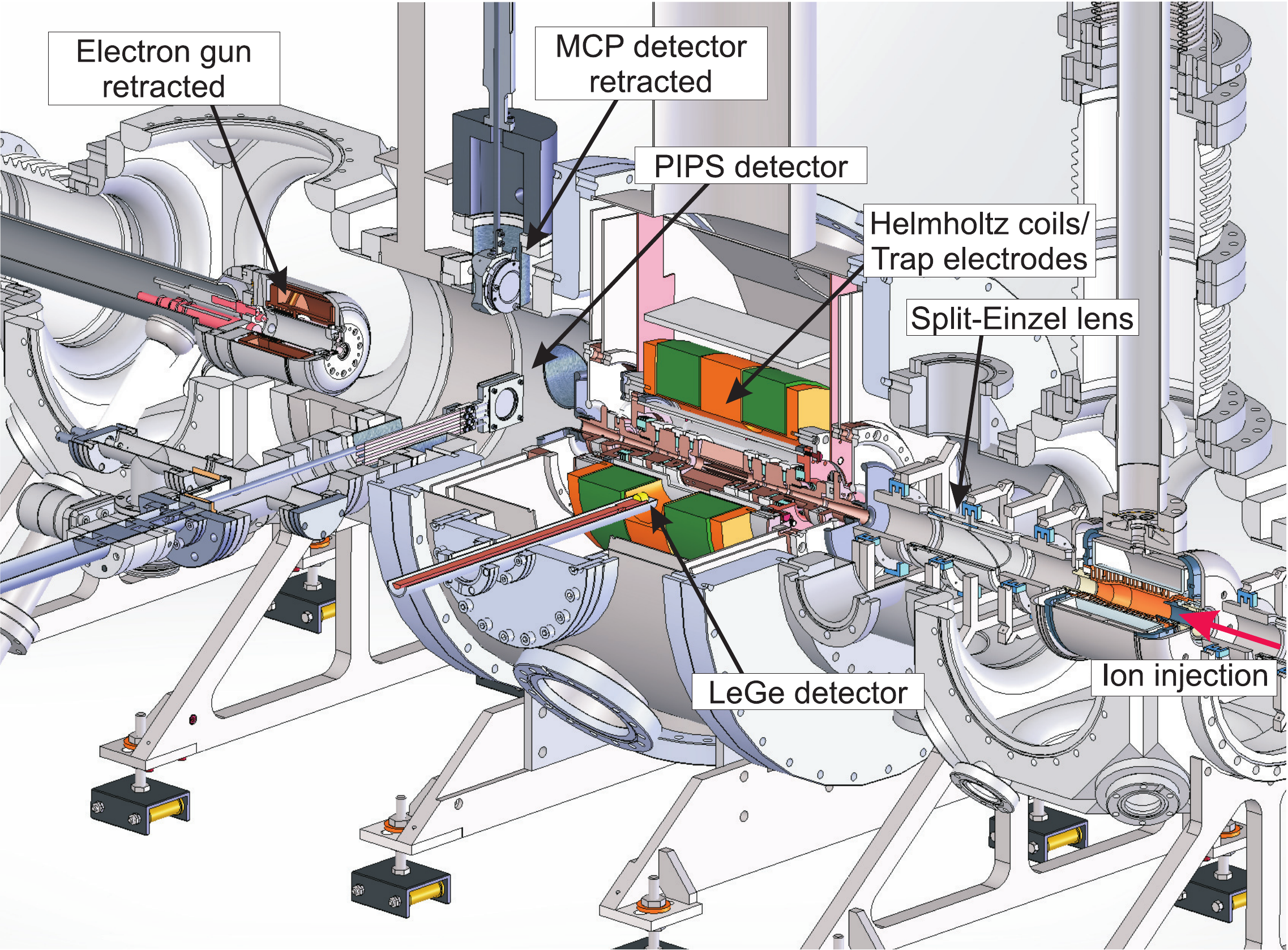}
	\caption{(color-online) Schematic rendering of the setup illustrating the trap center geometry as well as the position of LeGe detector (yellow cylinder), PIPS detector and retracted electron gun and MCP.}
	\label{fig:EBIT-schematic}
\end{figure}
%%%%%%%%%%%%%%%%%%%%%%%%%%%%%%%%%%%%%%%%%% First in-trap decay spectroscopy
\section{In-Trap Decay Spectroscopy}\label{firstITDS}
The feasibility of measuring electron-capture branching ratios with a Penning trap has been demonstrated by measuring the ECBR
of $^{124}$Cs within this work. This isotope was chosen due to its relatively large EC branch of 10.0(7)\%. During this measurement in-trap decay spectroscopy of $^{126}$Cs has been performed as well. Cesium-126 has been used to calibrate the efficiency of the detector. % and determine the ECBR of $^{124}$Cs. %The measured photopeak intensities were then used to
%determine the detection efficiency of the LeGe thus eliminating the geometrical acceptance of the detector. 
Both isotopes were produced as singly charged ions at the TRIUMF ISOL-type facility ISAC \cite{Dom02} by
bombarding a tantalum target with a 50\,$\mu$A proton beam at 500\,MeV delivered by the TRIUMF cyclotron.
Radioactive cesium atoms were surface-ionized and then passed through a high-resolution dipole magnet (resolving power $R\sim3000$) to reduce isobaric
contamination. Production yields were $1.3\cdot 10^7/$s for $^{124}$Cs, $2.5\cdot 10^5/$s for $^{124\textnormal{m}}$Cs, and $1.6\cdot 10^7/$s for
$^{126}$Cs measured at the ISAC-I yield station \cite{Kunz09}. At TITAN, the radioactive ions were cooled and bunched by the RFQ \cite{Smi06,Bru10b}. To identify the dominating isotope in the beam and to estimate the ion bunch intensity, 10 bunches were extracted from the RFQ at a rate of 10\,Hz and transported to a surface-barrier detector assembly in front of the MPET. This detector assembly consisted of an Al foil in front of a Passivated Implanted Planar Surface-barrier detector (Canberra PIPS detector \cite{Bru08}). The ions were implanted onto the Al foil. This detector is indicated as monitoring PIPS in Fig.~\ref{fig:TITAN-diagnostics}. The isotopes
$^{124}$Cs and $^{126}$Cs were identified by their half-lives. The half life of $^{124}$Cs was determined as 30.9(1)\,s which is in agreement with the literature value of 30.8(5)\,s \cite{Kat08}, however with a slightly improved uncertainty. Assuming a $\beta$-detection
efficiency of $(25\pm10)\%$ ion bunch intensities at 10\,Hz were estimated, with exponential fits to their decay times,
to be $(2.8\pm1.1)\cdot10^5$ ions/bunch for $^{124}$Cs and $(4.4\pm2.1)\cdot10^5$ ions/bunch
for $^{126}$Cs. The efficiency is the product of intrinsic efficiency and geometric
acceptance of the detector. The latter is the main contributor to the uncertainty of
the total efficiency due to the uncertainty of the distance of $(6\pm1)$\,mm between detector and
Al foil. The intrinsic detection efficiency was calculated by integrating the
$\beta$ spectrum \cite{Krane}. Based on this, an intrinsic detection efficiency of $\sim99\%$
was determined assuming that electrons below 250\,keV were stopped by the
20\,$\mu$m thick Al foil and the detector's dead layer, and thus not detected. The decay rate spectra were analyzed in order to investigate the possibility of isobaric Ba contamination. No significant contribution was found. 
%A background spectrum was then recorded with the LeGe detector and $^{126}$Cs
%being deposited in the RFQ, i.e., injecting radioactive isotopes \footnote{mention
%beam current here} into the RFQ without extracting them. The recorded spectrum
%is presented in Fig.\ref{fig:BGND} (red spectrum) along with a spectrum recorded prior to the delivery of radioactive cesium to TITAN (blue spectrum). Both spectra were normalized to one hour. The spectrum recorded with $^{126}$Cs being deposited inside the RFQ has a significantly higher count rate. Based on this observation it is
%evident that radioactive ions lost inside the RFQ significantly contribute to the photon background
%during an ECBR measurement. Photons originating from decays with sufficient energy
%are detected by the LeGe detector while low energy photons are attenuated by
%magnet bore and surrounding vacuum housing. 
%\begin{figure}
%        \centering
%	\includegraphics[width=.45\textwidth]{RFQ-dumped+BGND.eps}
%        \caption{\label{fig:BGND}Background measurements recorded with the LeGe
%detector. The red spectrum was recorded while $^{126}$Cs was deposited in the
%RFQ whereas the blue spectrum was recorded prior to the delivery of radioactive
%isotopes to TITAN.}
%\end{figure}
\subsection{Measurement Cycle and Data Acquisition} %%%%%%%%%%%%%%%%%%%%%%%%%%%%%%%%%% Measurement Cycle and Data Acquisition
During an ECBR measurement radioactive ions lost inside the RFQ and along the TITAN beam lines contribute to the photon background \cite{Bru11a}. In order to correct the in-trap decay spectrum for this background in-trap decay and background spectra were recorded successively. During the measurement period
$\Delta t_{meas}$, $\gamma$- and X-rays following radioactive decays were detected with
the LeGe detector. After this measurement, the trap was emptied by extracting the remaining ions. About $1$\,ms after each ion bunch was
extracted from the trap, a background measurement was started for $\Delta t_{BGND}$. Typically, storage and background-measurement time intervals of $\Delta t_{meas}=\Delta t_{BGND}=25$\,ms were used. A cycle with $\Delta t_{meas}=\Delta t_{BGND}=50$\,ms was also applied for a measurement period of one hour during the measurement of the calibrant $^{126}$Cs. 
%The full measurement cycle is shown in Fig.\,\ref{fig:cycleCs}. 
\begin{figure}
				\centering
							\includegraphics[width=.48\textwidth]{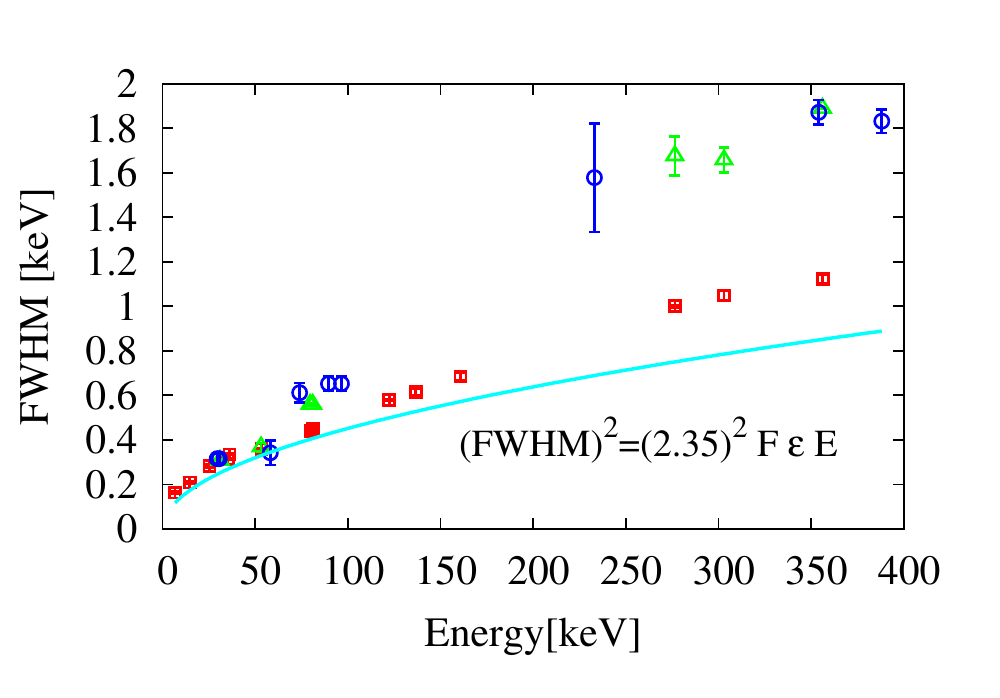}
							\caption{\label{fig:FWHM}(color online) Energy resolution (FWHM) of the LeGe detector in a zero $\left(\Box\right)$ and 2\,T magnetic field environment. The zero-field FWHM was measured using $^{57}$Co and $^{133}$Ba sources. The resolution at 2\,T was determined using $^{133}$Ba $\left(\bigcirc\right)$ as well as $^{124}$Cs and $^{126}$Cs calibrations lines $\left(\bigtriangleup\right)$. The solid line represents the resolution limit of a Ge detector determined by the Fano factor, $F$, and the energy, $\varepsilon$, required to create an electron-hole pair \cite{Knoll}.}
\end{figure}

The energy of the $\gamma$- and X-rays deposited in the crystal of the low-energy Ge detector was amplified by a transistor\-/reset amplifier. This amplifier provided an output voltage signal with a rising edge between 0 and 5 V, which was proportional to the deposited energy. The amplifier also provided a transistor-reset signal, which was used to disable the data acquisition while the voltage signal was reset to its initial value. Both the energy and transistor-reset signals were split and fed into two ORTEC DSPEC units, called 319 and 321. These two units were then gated to record data during the
time intervals $\Delta t_{meas}$ (DSPEC 321) or $\Delta t_{BGND}$ (DSPEC 319). 
%\begin{figure}
%        \centering
%                \includegraphics[width=.4\textwidth]{TITAN-EC-cycle}
%        \caption{\label{fig:cycleCs}Timing cycle used during the in-trap decay spectroscopy
%measurement of $^{124,126}$Cs.}
%\end{figure}

After the ions were extracted from the trap back into the beam line they were sent to the monitoring PIPS detector installed in front of the MPET \cite{Bru08} (see Fig.\,\ref{fig:TITAN-diagnostics}). The $\beta$-particle counting rate of this detector was then used to continuously 
monitor the intensity stability of the experiment \cite{Bru11a}.
\subsection{Influence of the Magnetic Field on the LeGe Detector} %%%%%%%%%%%%%%%%%%%%%%%%% Influence of the Magnetic Field on the LeGe Detector
The LeGe detector, which is installed on one of the EBIT's radial ports, was inserted between the two Helmholtz coils at a position close to the trap center. % At this position the magnetic field was estimated to be about 40\% of the field at the trap center with its orientation parallel to the surface of the crystal. 
During the experiment the magnetic field at the trap center was 5\,T resulting in a calculated residual field at the position of the LeGe detector of about 2\,T with its orientation parallel to the surface of the crystal. It was observed that this magnetic field strength broadens the photopeaks. Figure~\ref{fig:FWHM} shows the full width at half maximum (FWHM) of photopeaks recorded with the LeGe detector in a no-field and 2\,T field environment, respectively. No-field (0\,T) refers to a residual field dominated by the earth's magnetic field. No change of detection efficiency was observed within the uncertainty. This experimental finding necessitates further investigations.
\begin{figure*}
        \centering  							
        \includegraphics[width=.8\textwidth]{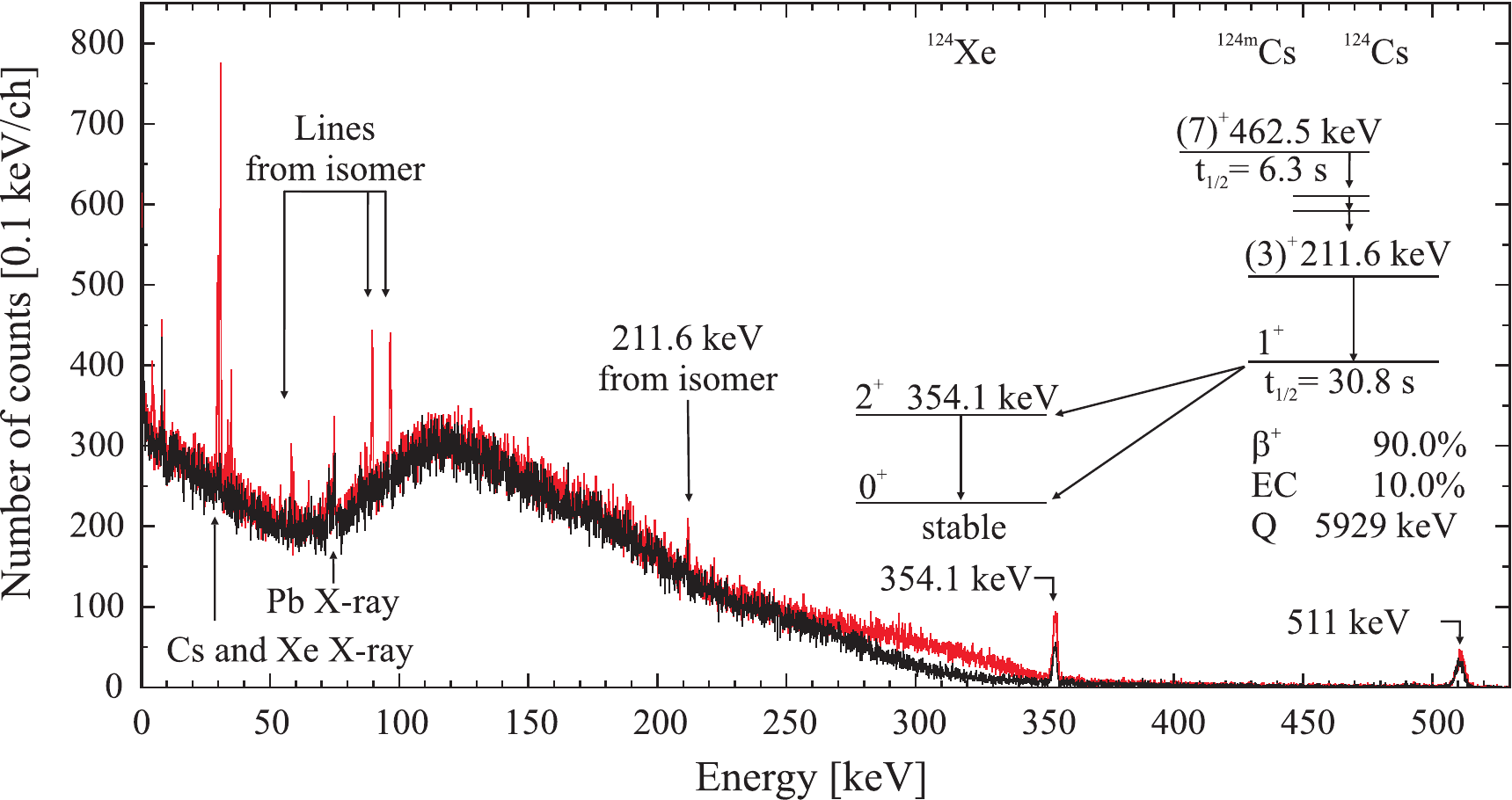}
        \caption{\label{fig:Cs124summed}(color online) Summed spectrum of the decay of $^{124}$Cs. The red spectrum was recorded with $^{124}$Cs stored in the trap whereas the black spectrum represents the background which was recorded with no ions stored. The measurement times of both spectra were identical. The simplified level scheme indicates the origin of the dominant photopeaks (from \cite{Kat08}, ECBR calculated with \cite{logft2001}).}
\end{figure*}
\subsection{Spectroscopic Measurements}%%%%%%%%%%%%%%%%%%%%%%%%%%%%%%%%%%%%%%%%%%%%%%%%%%%%%%%%%%%%%%%%%%%%%%%%%%%%%%%%%%%%%
The summed decay spectrum of the $^{124}$Cs EC-process is presented in Fig.~\ref{fig:Cs124summed}. The red spectrum was recorded while ions were stored in the ion trap and the black spectrum shows the background spectrum obtained after
each ion bunch was extracted from the trap. Indicated in these
spectra are the most prominent photopeaks and their origins. Clearly present are signatures of the isomer
$^{124\textnormal{m}}$Cs, i.e., the Cs $K$-shell X-ray lines of the electron conversion process and the photopeak at
211.6\,keV from the depopulation of the $3^+$ level in $^{124\textnormal{m}}$Cs. Also present in the spectrum are the Pb X-ray lines which appear in both background and in-trap decay spectra. It cannot be excluded that solder containing Pb was used to contact the Ge crystal, where Pb acts as an X-ray radiator. The Pb photopeaks do not overlap with any of the X-ray peaks of interest. In the in-trap decay spectrum a shoulder is observed between 250\,keV and 350\,keV. This shoulder occurs in the $^{124}$Cs and $^{126}$Cs spectra. 
%%%%%%%%%%%%%%%%%%%%%%%%% Jan10, 2013
%Radiative electron capture \cite{Gla56,Pac07} and synchrotron radiation can be excluded as source because the resulting photons have lower energies. 
%%%%%%%%%%%%%%%%%%%%%%%%
Calibration spectra with $^{133}$Ba show that the efficiency with which both DSPEC units record events degrades at energies above $\sim120$\,keV. In particular, a drop in efficiency is observed at 250\,keV for DSPEC 321 (background DSPEC) and 300\,keV for DSPEC 319 (in-trap measurement DSPEC). %resulting in the comparative lack of events in DSPEC 321 in the range 250\,keV to 350\,keV. 
This artifact does, however, not affect the analysis of the data presented in this publication. 

\begin{figure}
        \centering
                \includegraphics[width=.4\textwidth]{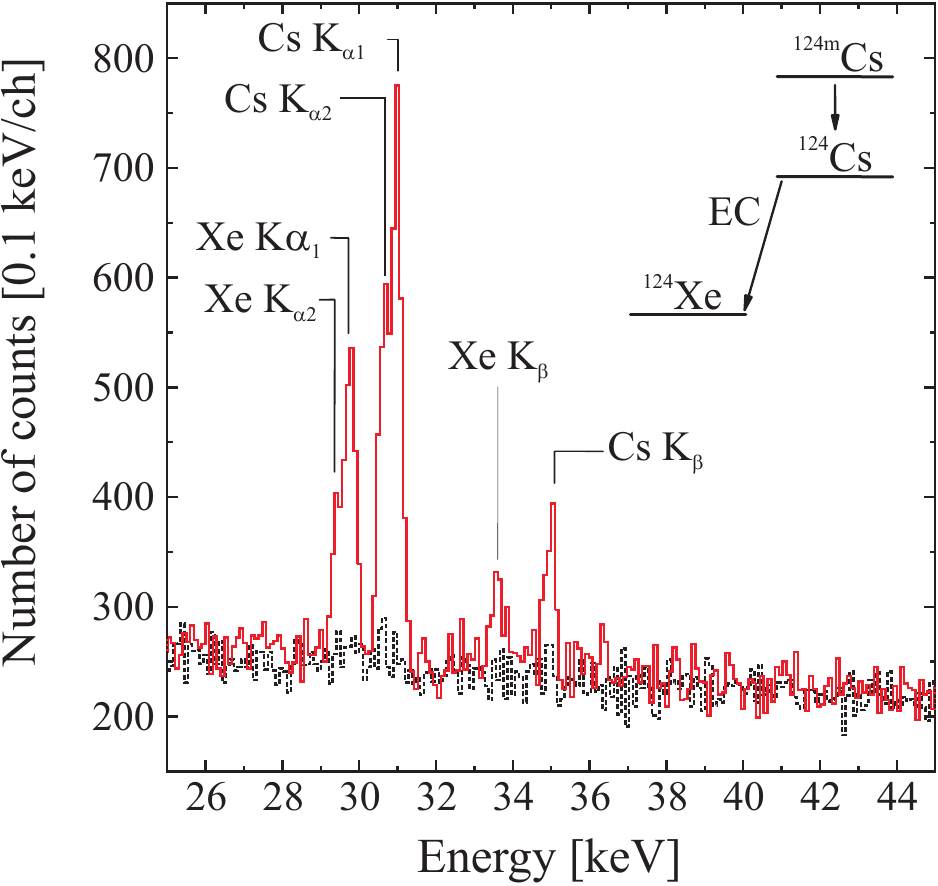}
        \caption{\label{fig:Cs124Xray}(color online) Energy region showing X-ray lines resulting from the decay of $^{124}$Cs. Dashed, black and solid, red spectrum were recorded during $\Delta t_{BGND}$ and $\Delta t_{meas}$, respectively. The measurement time was 6\,h.}
\end{figure}

A focused view of the X-ray region is presented in Fig.~\ref{fig:Cs124Xray} showing the advantage of the new in-trap decay
spectroscopy technique. The dominating $\beta$ decay did not contribute significant background to
the X-ray spectrum. Positrons reaching the LeGe detector would have resulted in
a drastically increased background in the in-trap decay spectrum (solid, red line). When ions are stored in the trap a clear xenon $K$-shell X-ray signature is visible originating from electron-capture decays of $^{124}$Cs. Also
present are the cesium $K$-shell X-ray lines that originate from the de-excitation
of $^{124\textnormal{m}}$Cs. Note that no $K$-shell X-ray peaks are observed in the background spectrum (dashed, black line). Comparing these two spectra one can conclude that no significant ion losses occurred radially and within the line of sight of the detector. %If ions were lost radially inside the trap a photopeak would be visible in both spectra. 

\begin{figure}
        \centering
                \includegraphics[width=.4\textwidth]{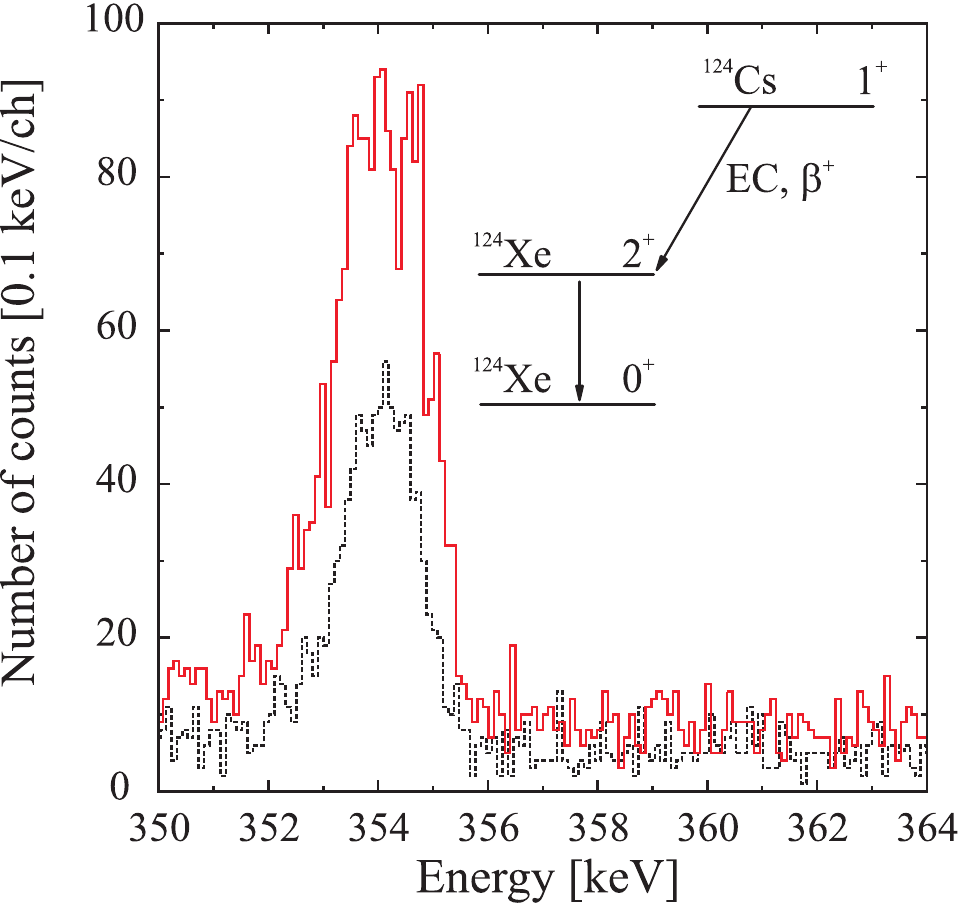}
        \caption{\label{fig:Cs124-354}(color online) 354\,keV photopeak originating from the $2^+$ level
of $^{124}$Xe. Dashed, black and solid, red spectrum were recorded during $\Delta t_{BGND}$ and $\Delta t_{meas}$, respectively. The spectrum is the sum of 6\,h of measurement.}
\end{figure}

The clear distinction between in-trap decay and background spectra worsens in the higher energy range. This is evident in the 211.6\,keV photopeak of the isomeric state and the 354\,keV photopeak originating
from the $2^+$ level in $^{124}$Xe. The latter peak is presented in
Fig.~\ref{fig:Cs124-354} with solid, red and dashed, black line being the in-trap decay and background
spectra, respectively. In this figure the photopeak is present in both spectra. In the background spectrum it originates from $^{124}$Cs ions that were lost along the beam line during injection and ejection as well inside the RFQ. The material surrounding the LeGe detector does not provide sufficient shielding. This agrees with previous observations where $^{126}$Cs was deposited in the RFQ and the 354\,keV peak was present in the LeGe detector spectrum \cite{Bru11a}. It is therefore essential to measure both in-trap decay and background spectra immediately after each other.

\begin{figure}
        \centering
                \includegraphics[width=.4\textwidth]{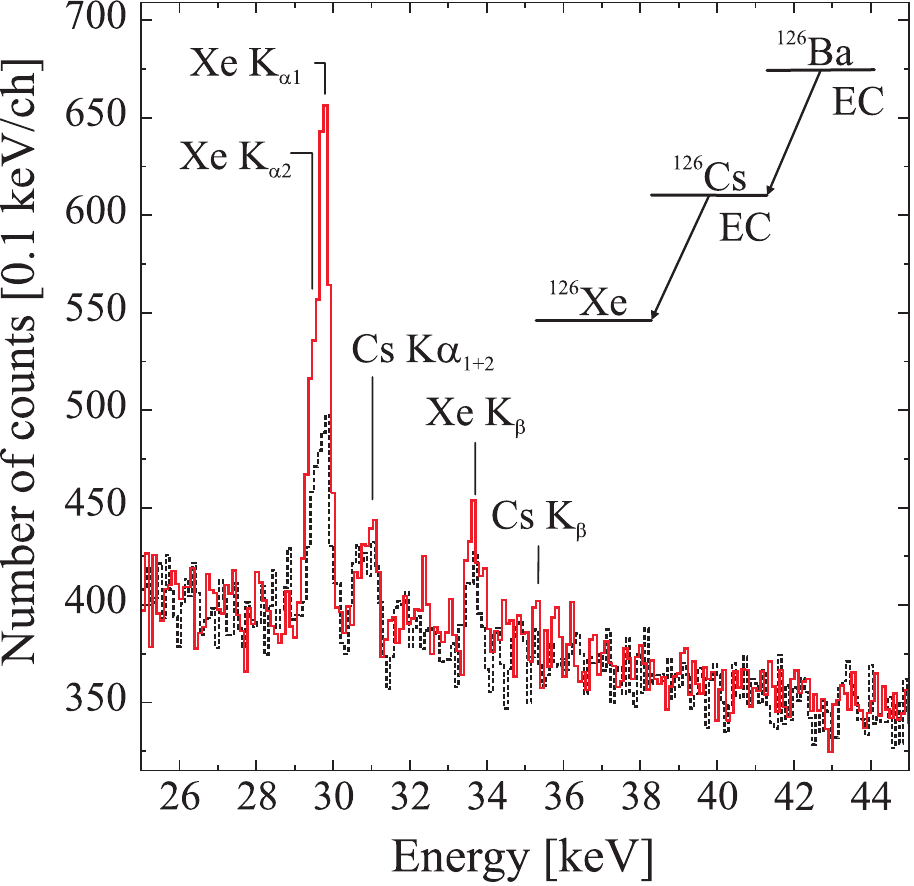}
        \caption{\label{fig:Cs126Xray}(color online) Energy region showing X-ray lines resulting from the decay of $^{126}$Cs. Dashed, black and solid, red spectrum were recorded during $\Delta t_{BGND}$ and $\Delta t_{meas}$, respectively. The spectra is the sum of 12\,h of measurement time.}
\end{figure}

\begin{figure}
        \centering
                \includegraphics[width=.4\textwidth]{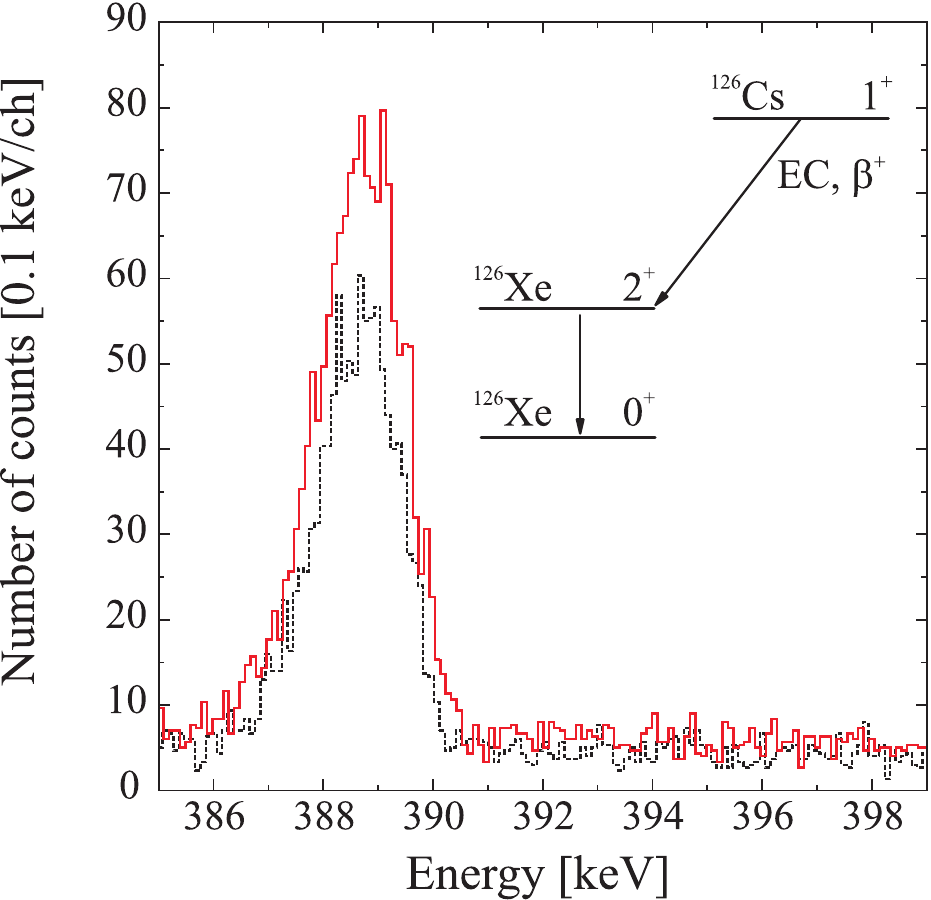}
        \caption{\label{fig:Cs126-388}(color online) 388\,keV photopeak originating from the $2^+$ level
of $^{126}$Xe. Dashed, black and solid, red spectrum were recorded during $\Delta t_{BGND}$ and $\Delta t_{meas}$, respectively. The total measurement time was 12\,h.}
\end{figure}

The $^{126}$Cs spectra that are used for calibration are presented in Fig.~\ref{fig:Cs126Xray} and Fig.~\ref{fig:Cs126-388} for X-ray and $\gamma$ region, respectively. During this measurement ions were lost inside the trap center and contributed to the X-ray background spectrum presented in Fig.~\ref{fig:Cs126Xray}. The 388\,keV photopeak also has a stronger contribution in the background spectrum as compared to the 354\,keV photopeak of $^{124}$Cs shown in Fig.~\ref{fig:Cs124-354}. Ion beam intensity as well as injection parameters were different for $^{124}$Cs and $^{126}$Cs. These effects need to be investigated in future developments of this technique in order to improve the signal-to-background ratio. 
\subsection{Data Analysis}
The electron capture branching ratio is defined as the number of decays by electron
captures $N_{EC}$ divided by the total number of decays $N$. The quantity
accessible in our experiment is the number of detected $K$-shell X-rays emitted
following an electron capture defined as $D_{K}^{meas}$. The total number of decays $N$ is extracted from the photopeak counts $D_{\gamma}^{meas}$ of the lowest $2^+$ state of the daughter nucleus of the $\beta$ branch. The photopeak intensity $I_{\gamma}$ was taken from literature \cite{Kat08} and the ECBR can then be written as
\begin{equation}
 BR(EC)=\frac{1}{\omega_{K}\cdot f_{K}}\left(\frac{\varepsilon_{int}^{det}(E_{\gamma})I_{\gamma}D_K^{meas}}{
\varepsilon_{int}^{det}(E_K)D_{\gamma}^{meas}}-\omega_K\ I_K^{CE}\right)\,,
\label{eq:ecbr}
\end{equation}
with $I_K^{CE}$ being the total intensity of the $K$-shell conversion electrons, $\omega_K$ being the fluorescence
yield, 
%i.e., the probability that the vacancy in the $K$-shell is closed while emitting an X-ray, 
and $f_K$ the probability that an EC event creates a vacancy in the $K$-shell. Since both, X-ray and $\gamma$ photons are recorded with the same detector the geometrical acceptance cancels out. The detection efficiency $\varepsilon_{int}^{det}(E)$ describes the intrinsic detection efficiency at X-ray and $\gamma$ energies $E_K$ and $E_{\gamma}$, respectively. The latter equation simplifies to
\begin{equation}
BR(EC)=\frac{I_K-I_{CE}}{\omega_K \cdot f_K},
\label{eq:BR}
\end{equation}
using the total $K$-shell intensity
\begin{equation}
I_K=\frac{\varepsilon_{int}^{det}(E_{\gamma})I_{\gamma}D_K^{meas}}{
\varepsilon_{int}^{det}(E_K)D_{\gamma}^{meas}}\,,
\label{eq:IK}
\end{equation}
and subtracting the $K$-shell X-ray intensity due to conversion electrons $I_{CE}=\omega_K I_K^{CE}=$\\$=0.888(5)$\cite{Sch96a}$\,\cdot\,0.010285(113)$\cite{Kat08}$=0.00913(11)$.
%%%%%%%%  old text
%The branching ratio is extracted from the measured counts $D_K^{meas}$ and
%$D_{\gamma}^{meas}$ in the X-ray and $\gamma$ photopeaks, respectively, the
%total intensity of the $\gamma$ transition $I_{\gamma}$, and the total intensity of
%K-shell conversion electrons $I_K^{CE}$. $\omega_K$ is the fluorescence yield,
%i.e., the probability that the vacancy in the $K$-shell is filled under the
%emission of an X-ray, and $f_K$ describes the probability that an EC leaves a
%vacancy in
%the $K$-shell. Since both, X-ray and $\gamma$ photons are recorded with the same
%detector the geometrical acceptance cancels out in this equation. $\varepsilon_
%{det}^{int}(E)$ describes the intrinsic detection efficiency at X-ray
%and $\gamma$ energies $E_K$ and $E_{\gamma}$, respectively. 
%%%%%%%%%%%%%%%%%%%%%%%%%%%
The detector efficiency term in the X-ray energy region cancels out if another isotope of the same element with a well-known X-ray intensity is measured under the same experimental conditions. This known X-ray intensity can be expressed as
\begin{equation}
 I_{K'}=\frac{D_{K'}^{meas}I_{\gamma'}\varepsilon_{int}^{det}\left(E_{\gamma'}\right)}{D_{\gamma'}^{meas}\varepsilon_{int}^{det}(E_{K'})}\,.
\end{equation} 
One can re-write Eq.~\ref{eq:IK} as
\begin{equation}
 I_K=I_{K'}\ \frac{D_K^{meas}\ D_{\gamma'}^{meas}\ I_{\gamma}\
\varepsilon_{int}^{det}(E_{\gamma})}{D_{K'}^{meas}\ D_{\gamma}^{meas}\ I_{\gamma'}\
\varepsilon_{int}^{det}(E_{\gamma'})}\,,
\label{eq:IK-new}
\end{equation}
if the X-ray energies are identical, as for isotopes of the same element. 
%The isotope-dependent X-ray energy shift is negligible compared with the energy resolution of the detector. 
Therefore, only the ratio $\varepsilon_{int}^{det}(E_{\gamma})/\varepsilon_{int}^{det}(E_{\gamma'})$ needs to be
known. For photon energies above $\sim120$\,keV the detector
response is well known and the self-shielding of the calibration sources is less of
a concern than in the X-ray region. Typically, $E_{\gamma}$ and $E_{\gamma'}$ are above
120\,keV. The ECBR of $^{124}$Cs was determined by applying the described method
using $^{126}$Cs for calibration.
\subsection{Results}
The photopeak counts $D_{K}^{meas}$ and $D_{\gamma}^{meas}$ of the $^{124}$Cs, $^{126}$Cs and
calibration spectra have been determined by the method of maximum likelihood. In the fit of X-ray spectra the relative $K$-shell intensities and energies were constrained to their known values \cite{XRDB}. %The fits were performed using the same method used at the University of Washington to determine the ECBR of $^{100}$Tc \cite{Sju09}, the intermediate nucleus of the $\beta\beta$ candidate $^{100}$Mo. 
%Comparison with Monte-Carlo simulations shows that both DSPEC units attenuated energy signals of photon energies above $\sim120$\,keV. At photopeak energies of $\sim300$\,keV this truncation amounts to more than two orders of magnitude. This attenuation was appropriately taken care of, but at a cost of a large systematic error.
%During the analysis of the Cs spectra and in comparison with PENELOPE2008 \cite{Pen08} efficiency simulations it was discovered that both DSPEC units attenuated energy signals of photon energies above $\sim120$\,keV. At photopeak energies of $\sim300$\,keV this truncation amounts to more than two orders of magnitude. Furthermore, it was found that this truncation differs for both DSPEC units. This difference was accounted for by independently determining $\varepsilon(E_{\gamma})$ and $\varepsilon(E_{\gamma'})$ for each unit. This is reflected in the large systematic uncertainty of the $^{124}$Cs ECBR.

The measurement periods $\Delta t_{meas}=\Delta t_{BGND}$ of
25\,ms and 50\,ms were short compared to the half-lives of 30.8(5) s and 1.64(2) min for $^{124}$Cs and $^{126}$Cs, respectively.
%compared to the half-lives of $^{124}$Cs and $^{126}$Cs of 30.8(5)\,s and 1.64(2)\,min, respectively. 
Therefore, the decay rate was considered constant throughout both measurement periods. This assumption is valid to a level of the order of $10^{-4}$.    % and was verified in simulations.
Based on this assumption, the peak intensity of the background measurement recorded during $\Delta t_{BGND}$ was subtracted from the peak
intensity measured during $\Delta t_{meas}$. The measurement was divided into time slices. Within each time slice all settings were kept constant. Cesium-126 was measured in five time slices for a total of 12 hours while $^{124}$Cs was measured in two time slices for a total of 6 hours. The resulting $^{124}$Cs and $^{126}$Cs photopeak intensities are presented in Table \ref{tab:124Cs} and Table \ref{tab:126Cs}, respectively.
\begin{table}
\caption{\label{tab:124Cs}Counts in the photopeaks of $^{124}$Cs. The uncertainty is determined by the fit and listed in brackets. 0(10) is unphysical but arises from the fitting routine. For consistency reasons the value is presented this way. Two time slices (TS) of 4 and 2 hours were measured. The spectra within each time slice were summed prior to the analysis. The fourth column lists the $\gamma$-peak counts corrected for the different DSPEC inefficiencies.  The last column indicates whether the data was recorded during $\Delta t_{BGND}$ (BGND) or $\Delta t_{meas}$ (storage).}
% \begin{ruledtabular}
\centering
 \begin{tabular}{lcccr}
  TS&X-ray peak&$\gamma$ peak&corr. $\gamma$ peak&\\
	$[\textnormal{h}]$&Xe $K_{\alpha}$\&$K_{\beta}$&354.1\,keV&354.1\,keV&\\
  \hline\noalign{\smallskip}
  \multirow{2}{*}{4}&94(43)&549(28)&837(158)&BGND\\
  &1074(66)&1152(40)& &storage\\
  \hline\noalign{\smallskip}
  \multirow{2}{*}{2}&0(10)&285(19)&434(84)&BGND\\
  &490(42)&534(27)&&storage\\
\noalign{\smallskip}\hline
 \end{tabular}
% \end{ruledtabular}
\end{table}
%%%%%%%%%%%%%%%%%%%%%%%%%%%%%%%%%%%%%%%%%%%%%%%%%%%%%%%%%%%%%%%%%%%%%%%%%%%%
\begin{table}
\caption{\label{tab:126Cs}Counts in the photopeaks of $^{126}$Cs. The values are listed in the same way as those of $^{124}$Cs presented in Table\,\ref{tab:124Cs}.}
% \begin{ruledtabular}
\centering
 \begin{tabular}{lcccr}
  TS&X-ray peak&$\gamma$ peak&corr. $\gamma$ peak&\\
	$[\textnormal{h}]$&Xe $K_{\alpha}$\&$K_{\beta}$&388\,keV&354.1\,keV&\\
  \hline\noalign{\smallskip}
  \multirow{2}{*}{3}&399.67(67)&704(31)&843(125)&BGND\\
  &1218.91(69)&1038(42)&&storage\\
  \hline\noalign{\smallskip}
  \multirow{2}{*}{1}&164(45)&382(22)&271(43)&BGND\\
  &591(53)&479(26)&&storage\\
	  \hline\noalign{\smallskip}
  \multirow{2}{*}{2}&494.89(62)&730(30)&874(128)&BGND\\
  &864.67(68)&896(34)&&storage\\
	  \hline\noalign{\smallskip}
  \multirow{2}{*}{3}&513.75(62)&667(29)&798(118)&BGND\\
  &1055.45(74)&947(35)&&storage\\
	  \hline\noalign{\smallskip}
  \multirow{2}{*}{3}&480.26(58)&603(27)&722(107)&BGND\\
  &963.81(68)&799(32)&&storage\\
\noalign{\smallskip}\hline
 \end{tabular}
% \end{ruledtabular}
\end{table}
%%%%%%%%%%%%%%%%%%%%%%%%%%%%%%%%%% New Part on fucked up DAQ
After the experiment was performed it was discovered that the DSPEC units truncate the energy signal, i.e., the number of recorded counts in the spectrum does not correspond to the number of expected counts. Comparison with Monte-Carlo simulations shows that both DSPEC units attenuated energy signals of photon energies above $\sim120$\,keV. At photopeak energies of $\sim300$\,keV this truncation amounts to more than two orders of magnitude. During the experiment $^{133}$Ba calibration spectra were recorded to identify possible drifts in the data acquisition. However, the number of counts in these spectra are low hence it is difficult to use them for calibration.
The ratio of photopeak counts recorded with DSPECs 319 and 321 was calculated for all peaks in the $^{133}$Ba spectrum. The ratio in the X-ray region agrees with unity within uncertainty. However, the ratio at photopeak energies 276.3 keV, 302.9 keV, and 356.0 keV is energy dependent following a linear curve. This curve determines how efficient DSPEC 319 records photons at the energies of 354\,keV ($^{124}$Cs) and 388\,keV ($^{126}$Cs) with respect to DSPEC 321. The extrapolated ratio was then used to correct the photopeaks recorded with DSPEC 321 at 354\,keV and 388\,keV. The corrected values as well as all extracted photopeak counts are listed in Table\,\ref{tab:124Cs} and Table\,\ref{tab:126Cs} for $^{124}$Cs and $^{126}$Cs, respectively. The corrected counts in the background spectrum were then subtracted from the counts of the spectroscopy spectrum and the ratios of $D_{K}^{meas}/D_{\gamma}^{meas}$ and $D_{K'}^{meas}/D_{\gamma'}^{meas}$ were determined. These ratios were calculated independently for each time slice. The weighted average of this ratio was then calculated for each isotope. The ratio $\frac{\varepsilon_{int}^{det}(E_{\gamma})}{\varepsilon_{int}^{det}(E_{\gamma'})}$ was determined with the $^{133}$Ba calibration spectrum recorded with DSPEC 319. The $^{133}$Ba photopeaks at energies of 276.3\,keV, 302.9\,keV and 356.0\,keV \cite{Kha11} were used to extract the ratio of efficiencies at 354\,keV and 388\,keV.
%%%%%%%%%%%%%%%%%%%%%%%%%%%%%%%%%% End of new part

The ECBR of $^{124}$Cs was calculated according to Eq.\,\ref{eq:BR} and Eq.\,\ref{eq:IK-new} to be $(22.7 \pm 8.9(\textnormal{stat.}) \pm 14.8(\textnormal{syst.}))\%$ and is based on the measured ratios $\frac{D_{K}^{meas}}{D_{\gamma}^{meas}}$ ($^{124}$Cs), $\frac{D_{K'}^{meas}}{D_{\gamma'}^{meas}}$ ($^{126}$Cs), and $\frac{\varepsilon_{int}^{det}(E_{\gamma})}{\varepsilon_{int}^{det}(E_{\gamma'})}$, $f_K = 0.841(60)$ \cite{Kat08}, $\omega_K = 0.888(5)$ \cite{Sch96a}, $I_{CE}=0.009133(113)$, and $I_{K'}=0.146(2)$ \cite{Kat02}. Our value agrees with the literature value of 10.0(7)\% \cite{Kat08,logft2001}. However, the truncated detection efficiency of the DSPEC units reduced the counts in the $\gamma$ photopeak that was used for normalization. The inefficiency in detection of these photons artificially increases the ECBR and is accounted for in the large systematic uncertainty. This systematic uncertainty arises from the different efficiencies of DSPEC 319 and 321. Nevertheless, the presented measurement proves that these decay-spectroscopy experiments are feasible.

The use of detected $\beta$ particles at the PIPS detector mounted at one side of the trap for normalization has been investigated. In detailed SIMION \cite{Dahl:2000lr} simulations it was found that the number of electrons or positrons reaching the PIPS detector strongly depends on various parameters \cite{Bru11,Bru11a}. %The largest impact has the position of the decaying ion, i.e., ion could size and ion distribution, inside the trap. 
Among those, the position of the decaying ions (i.e. the ion cloud size and density distribution inside the trap) has the largest impact. 
The probability of a $\beta$ particle reaching the PIPS detector also depends on the $Q$ value of the decay. Since the trap is cryogenic it is almost impossible to place a calibration source inside the trap center and map the $\beta$ detection efficiency for various source locations. Hence the concept of using $\beta$ particles for normalization purposes has been discarded.

The X-ray photopeak areas of $^{124}$Cs and $^{126}$Cs, and the 89.5\,keV and 96.6\,keV lines of
$^{124\textnormal{m}}$Cs were used to estimate the number of ions that were injected into the trap.
%These peaks were chosen to determine the ion bunch intensity.
% since in the region below $\sim120$\,keV, the signal was not attenuated by the DSPEC units. 
Only decays of stored ions were considered, i.e., the peak area recorded during the background measurement was
subtracted from the decay measurement. The detection efficiency was extracted from PENELOPE simulations of the experimental setup. The extracted intensities are presented in Table\,\ref{tab:intensities}. The intensities of $^{126}$Cs were extracted for each time slice separately throughout the experiment. They varied over time due to slight changes in beam tune and delivered isotope intensity. The RFQ cycle that was applied during these measurements differs from the 10\,Hz rate presented in Section\,\ref{firstITDS}. Furthermore, before being captured in the Penning trap the ion bunches had to pass two additional electrostatic $45^{\circ}$ bender where ions could have been lost. This could explain the lower ion bunch intensities listed in Table\,\ref{tab:intensities} compared to the PIPS measurement presented earlier.
\begin{table}
	\caption{\label{tab:intensities}Ion bunch intensities derived from the measured $^{124\textnormal{g,m},126}$Cs photopeak
intensities and the duration of each time slice (TS). The value marked with $^{\dagger}$ was measured with $\Delta t_{meas}=\Delta t_{BGND}=50$\,ms. All other measurements were performed with 25\,ms.}
%\begin{ruledtabular}
	\centering
		\begin{tabular}{rccc}
			Isotope&TS&TS $[\textnormal{h}]$&Ion bunch intensity\\
			  \hline\noalign{\smallskip}
			$^{124\textnormal{m}}$Cs&&\multirow{2}{*}{6}&$4.48(63)\cdot10^3$\\
			$^{124\textnormal{g}}$Cs&&&$1.34(13)\cdot10^5$\\
			$^{126}$Cs&1&3&$2.65(32)\cdot10^5\ ^{\dagger}$\\
			$^{126}$Cs&2&1&$2.54(54)\cdot10^5$\\
			$^{126}$Cs&3&2&$1.85(46)\cdot10^5$\\
			$^{126}$Cs&4&3&$1.80(32)\cdot10^5$\\
			$^{126}$Cs&5&3&$1.61(30)\cdot10^5$\\
			\noalign{\smallskip}\hline
		\end{tabular}
%	\end{ruledtabular}
\end{table}
\section{Future ECBR measurement} 
After the ECBR measurement described in this work the detection setup has been upgraded. The single LeGe detector has been replaced by seven custom-made lithium-drifted Silicon (Si(Li)) detectors installed at all
available view ports of the trap. This increases the geometrical acceptance from $\sim0.1\%$ \cite{Bru11c} to $\sim2.1$\%. The detectors are placed outside the vacuum vessel behind thin Be windows. This increased distance to the trap center reduces the magnetic field at the position of the crystal to less than 1\,T.  To further enhance the
sensitivity of the setup towards smaller ECBRs passive low-activity Copper-Lead sandwich shields have been added to the Si(Li) detectors. Moreover, the addition of BGO crystals as active Compton-suppression shielding has been studied, and a prototype is being
constructed. The data acquisition system has been upgraded by replacing the DSPECs with sampling analog-to-digital converters (Struck SIS3302) \cite{Len12}, effectively eliminating any rate or energy dependent biases. These upgrades significantly reduce the background while increasing acceptance and mitigate the presented issues with the DSPEC DAQ.  %In its final configuration the experiment will use 
\section{Conclusion}
A new technique has been developed at the TITAN facility to measure the ground-state transition strength of intermediate nuclei in
$\beta\beta$ decays utilizing an open-access Penning trap. The feasibility and power of
this in-trap decay spectroscopy technique have been demonstrated and various systematic
studies have been performed. Within these studies an electron-capture decay of ions
stored in a Penning trap has been observed for the first time \cite{Bru11c,Ett09}. During the ECBR measurement of $^{124}$Cs it was observed
that ion losses along the beam line also contribute to the recorded photon spectrum,
dominantly for photopeak energies above $\sim120$\,keV. A measurement cycle
has been developed and applied to record background spectra immediately after the
ions were extracted from the trap. 
%This step is crucial if a $\gamma$-ray photopeak is used for normalization. 
With this technique and the upgraded experimental setup the determination of ECBRs as low as $10^{-5}$ is within reach.
%
%%%%%%%%%%%%%%%%%%%%%%%%%%%%%%%%%%% Acknowledgments
\section{Acknowledgment}
TRIUMF receives federal funding via a contribution agreement with the National Research
Council of Canada (NRC). This work was partly funded by NSERC and by the Deutsche Forschungsgemeinschaft (DFG) under
grant FR 601/3-1. T.B. acknowledges
support by evangelisches Studienwerk e.V. Villigst, A.G. acknowledges support from NSERC
PGS-M program, V.V.S from the Studienstiftung des Deutschen Volkes and S.E. from the Vanier CGS program. We acknowledge the great support of TRIUMF and ISAC operations. We thank I. Ostrovskiy for fruitful discussions on data analysis and statistics. Also thanked are C. Pearson, S. Williams, S. Daviel, and P.-A. Amaudruz for their help with the DAQ system. We would especially like to thank TITAN's technologist M. Good for his help in setting up this experiment.
 \bibliographystyle{epjc}
 \bibliography{titan-ec}
%\bibliographystyle{}
%\bibliography{titan-ec}% Produces the bibliography via BibTeX.
%\bibliographystyle{ieeetr}
\end{document}